# Laser floating zone growth of $SrVO_3$ single crystals

Tanya Berry[a,b], Shannon Bernier[a,b], Gudrun Auffermann[c], Tyrel M. McQueen[a,b,d], and W. Adam Phelan[†,a,b*]

[a] Department of Chemistry, The Johns Hopkins University, Baltimore, MD 21218, USA

[b] Institute for Quantum Matter, Department of Physics and Astronomy, The Johns Hopkins University, Baltimore, MD 21218, USA

[c] Max-Planck Institute for Chemical Physics of Solids, Nöthnitzer Str. 40, D-01187 Dresden, Germany

[d] Department of Materials Science and Engineering, Johns Hopkins University, Baltimore, MD 21218, USA

† Present address: Los Alamos National Laboratory, Los Alamos, NM 87545

*corresponding author

**Abstract:**

The perovskite $SrVO_3$ is of interest for a variety of applications due to its simple metallic character and stability in reducing environments. Here we report the preparation of single-crystal $SrVO_3$ using the laser floating zone technique. Laue diffraction implies single domains ~30 mm in length. The stoichiometry of optimized crystals was found to be $Sr_{0.985(7)}VO_{2.91(4)}$ using inductively coupled plasma optical emission spectrometry and neutron powder diffraction analysis, with compositions adjustable depending on the crystal pulling rate. Heat capacity measurements from T = 2 – 300 K show variations with composition, attributable to a combination of impurity scattering and changes in phonon dynamics. Our results demonstrate the utility of the laser floating zone technique in preparing a range of materials, and our advances with $SrVO_3$ may help lead to applications including catalysis, transparent conducting oxides, thermionic emitters, and other electronic devices.

## Introduction:

The perovskite family is arguably one of the most intensely studied material classes in terms of highly functional materials. The basic cubic perovskite $ABX_3$ crystal structure was first studied in the mid-1920s by V. M. Goldschmidt, and subsequently lower-symmetry variants were discovered in the late 1940s.[1,2] In an ideal cubic perovskite, the A cation is 12-fold coordinated while the B cation is 6-fold coordinated to the X anion.[3,4] The $BX_6$ octahedra form a 3D corner-sharing network with cavities occupied by the A cations.[2,4] Historically, oxide perovskites like $SrVO_3$ have been very well-studied, in part because the -2 oxidation state of oxygen is compatible with a wide variety of cation oxidation states. This, combined with oxygen's small radius fitting well inside the unit cell, makes oxide perovskites more abundant than halides, nitrides, or other chalcogenides.[4] Oxide perovskites have been used for such varied purposes as sensing, catalysis, and high-temperature superconductors.[4] $SrVO_3$ specifically, is of interest as a transparent conducting oxide[5,6], catalyst[7,8], and thermionic electron emitter. [9-11]

In $SrVO_3$, vanadium is in a +4 oxidation state with $d^n$ itinerant electrons, with the oxygen 2p band filled below the vanadium 3d bands.[9] Oxides such as $SrVO_3$ are known to change from metallic to insulating due not only to alternations electron count but also the degree of overlap between neighboring atoms. With a 180° V-O-V bond angle and 1/6 filling of the $t_{2g}$ manifold, ideal $SrVO_3$ is expected, and experimentally observed, to be metallic, with moderate electron correlations.[12] However, changes in electron count and V-O-V bond angle by substitution are known to drive metal-insulator transitions in thin film $SrVO_3$ and in doped or substituted bulk $SrVO_3$.[12]

Single crystals of $SrVO_3$ in this work were produced using a tilting laser diode floating zone (tiltLDFZ). In this device five laser diodes are used to heat the interface between two solid rods (the "seed" and "feed" rods), producing a molten zone from which the desired crystal is grown. The technique offers enhanced control over the temperature and stability of the molten zone compared to other floating zone techniques.[13] It has been used to grow single crystals of a wide range of materials that pose a variety of synthetic challenges, including incongruent melters, crystals prone to undesired twinning, materials with high-vapor pressure precursors, and high entropy pyrochlores.[14-20]. In these cases, the improved control over the molten zone allowed the various synthetic challenges to be addressed.

The characterization of the composition of materials is equally important. In the broader materials synthesis community, chemical compositions are often determined using SEM-EDX techniques that utilize a tear drop model which is prone to large error. He we combine techniques with complementary sensitivities to assign overall composition. For example, Sr:V ratios are best studied using X-ray and ICP-OES techniques because of their high atomic numbers, whereas Sr:O ratios are best characterized using neutron diffraction since O is very visible in neutron diffraction.

Much work on $SrVO_3$ to date has been in the thin-film community, which has achieved great success in producing highly pure $SrVO_3$ thin films for functional purposes.[6,16] Here, we introduce a synthetic technique to produce single-crystalline $SrVO_3$ with controlled defect chemistry as confirmed by XRD, ICP-OES, and neutron diffraction. We also present measurements of heat capacity in the bulk crystals, showing variations with changes in defect chemistry, and compare the resistivity of single crystalline $SrVO_3$ to previous literature reports, concluding that subtle variations in crystal chemistry profoundly affect the low temperature physical properties.

## Experimental:

### Synthesis of polycrystalline $SrVO_3$

Single crystal samples of $SrVO_3$ were grown from polycrystalline $SrVO_3$/$Sr_2V_2O_7$ compacts. These compacts were prepared in two steps: first, $Sr_2V_2O_7$ was synthesized from thoroughly mixed batches of dried $SrCO_3$ (Sigma-Aldrich, 99.9% purity) and $V_2O_5$ (from NOAH Technologies, 99.99% purity), ~10 g each, in the molar ratio of 2:1, which were heated to 1000°C for 24 hours in a box furnace in an open alumina crucible as homogenously mixed powders. Four subsequent regrinds and reheatings were performed until it was observed via X-ray powder diffraction that only $Sr_2V_2O_7$ was present. Second, powders of $Sr_2V_2O_7$ were reduced at 1000°C for 24 hr in a mullite tube furnace under forming gas (Ar:$H_2$ 95%:5%) with a flow rate 60 mL/min to form $SrVO_3$. Again, subsequent regrinds and 2 reheatings under flowing forming gas were performed until it was observed via Rietveld refinement of X-ray powder diffraction data that the mixture was *ca*. 85% $SrVO_3$ and 15% $Sr_2V_2O_7$.

### Synthesis of $SrVO_3$ single crystals

Sintered rods approximately 4 mm in diameter and 80 mm in length were prepared by compacting and isostatically pressing the 85% $SrVO_3$/15 % $Sr_2V_2O_7$ powder into cylindrical rods. These rods were heated to T = 1200°C in a vacuum furnace for 8 hours under a $10^{-5}$ torr dynamic vacuum. Once sintered, the seed and feed rods were attached to the lower and upper shafts, respectively, of the tiltLDFZ furnace (Crystal Systems Inc. FD-FZ-5−200-VPO-PC) with 5×200 W GaAs lasers (976 nm). First, the seed rod was melted and then the molten zone was formed by bringing the feed rod into contact with the seed rod. During the growth, the seed and feed rods were counter-rotated at 10 rpm, and the crystal was formed by moving the seed and feed rods downward (molten zone fixed) at rates of 3 and 2, 6 and 4, and 9.5 and 6 mm/hr, respectively; these will be referred to as “2 mm/hr”, “4 mm/hr”, and “6 mm/hr” in the following sections. In order to maintain the reduced $V^{+4}$ oxidation state of $SrVO_3$, all growths were performed under 1 atm forming gas (in which $SrVO_3$ is a congruent melter) with a flow rate of 1 L/min.

### X-Ray Diffraction

Powder X-ray diffraction patterns were collected on a Bruker D8 Focus diffractometer with a LynxEye detector using Cu Kα radiation. Lattice parameters and Rietveld refinements were performed using Topas 4.2 (Bruker). A silicon standard was utilized to determine the lattice parameter among the growths. This is necessary because, in cubic materials there is only one lattice parameter which can convolute with sample displacement and diffractometer zero errors.[21]

Single crystal X-ray diffraction data were collected using a SuperNova diffractometer equipped with an Atlas detector and a Mo Kα source. The cuboid crystal, cut from a larger crystal piece, was mounted with Paratone-N oil. Data was analyzed and reduced using the CrysAlisPro software suite, version 1.171.36.32 (2013), Agilent Technologies. Initial structural models were developed using SIR92 and refinements of this model were done using SHELXL-97 (WinGX version, release 97-2).[22-23]

Crystal alignments were done using back-reflection Laue diffraction with a MultiWire Laboratories MWL110 detector. The X-ray beam about 1 mm in diameter was utilized.

**Neutron Diffraction**

Neutron powder diffraction was performed at the POWGEN beamline at the Spallation Neutron Source at Oak Ridge National Laboratory. The data were collected at T = 300 K and 10 K with 50-minute data collection times per temperature using wavelengths ranging from 0.27 – 1.33 Å (Q-range *ca.* 0.75 to 42 $Å^{-1}$) with wavelength center ~ 0.8 Å in the POWGEN Automatic Changer. The powder samples were loaded in vanadium cans of 8 mm inner diameter consisting of a copper gasket and aluminum lid and sealed under an argon atmosphere. Data were modelled starting with the reported cubic structure Rietveld refinement with GSASII[24] from 1.6 to 23.5 $Å^{-1}$ in order to extract the Sr:O ratio.

**Chemical Analysis**

Inductively coupled plasma-optical emission spectroscopy (ICP-OES) analysis was performed using an Agilent ICP-OES 5100 SVDV. Three accurately weighed-in samples (about 5 mg) were digested in 2 mL aqua regia by heating. Each solution was transferred into a volumetric flask (50 mL) and filled up with ultrapure water. Afterwards, these solutions were measured against a matrix matched calibration series to determine the Sr and V content and to check the solutions with regard to trace amounts detectable via ICP-OES such as alkaline earth and transition metals.

**Physical Properties**

Heat capacity and resistivity measurements were collected from single crystals of $SrVO_3$ using a Quantum Design Physical Properties Measurement System. Heat capacity data was collected from T = 2 – 300 K under applied fields of $\mu_0H$ = 0.1 T using the semi-adiabatic pulse technique with a 1% temperature rise and three repetitions at each temperature. The resistivity was measured from T = 3 – 300 K using the four-probe technique. The leads were made out of Pt wire and the contacts were made using Dupont 4922N Ag paste. The Pt lead distance was 0.38 mm. The sample thickness was 1.2 mm.

## Results and Discussion:

**Synthesis**

Although previously reported synthetic techniques use the same starting materials, prior reports did not first produce $Sr_2V_2O_7$ and, instead, used $SrCO_3$ and $V_2O_5$ directly in the rods used for the single crystal growth.[25] This has a number of disadvantages, including requiring use of a binding agent to help form the rods, the release of gaseous carbon dioxide leading to porosity of the heated rods, and the significantly different melting points of $SrCO_3$/SrO and $V_2O_5$ which can lead to inhomogeneities throughout the grown crystal. These methods result in various thermodynamically stable secondary phases including $Sr_3V_2O_7$, $Sr_4V_3O_{10}$, $Sr_5V_4O_{13}$, and $Sr_6V_5O_{16}$.[25] In this work, we avoid these issues by forming rods of majority $SrVO_3$ (85% purity). The use of forming gas in the LDFZ method allows for the reduction of the remaining $Sr_2V_2O_7$ to $SrVO_3$, enabling straight-forward growth conditions, in which synthetic conditions can be systematically varied.

Figure 1a shows a single crystal of $SrVO_3$ produced using our synthetic process at a rate of 2 mm/hr along with a Laue image showing sharp reflections with the growth direction along [110]. In contrast to previous synthetic methods, no binding agent was incorporated in the process of pressing rods[25] which minimizes the number of void defects that negatively affect the density and homogeneity of a crystal, and allows for the formation of large, dense, single crystalline domains. The lattice parameters of the samples synthesized here were found to show a linear increase with growth rate, Figure 1b, with the slowest growth rates most closely matching lattice parameters reported in the literature.

**Structure**

A combination of single crystal X-ray diffraction and powder neutron diffraction was used to determine the structural details of the as-prepared $SrVO_3$. Modeling of single crystal x-ray diffraction data collected at room temperature, Table 1, is in good agreement with the structures already reported in the literature. Site occupancies were found to be equal to unity within error, and thus fixed at unity in the final refinement. The refinement converged to very good agreement with the data, $R_1$ = 0.77% and $wR_2$ = 2.23%, with physically reasonable atomic displacement parameters.

Neutron diffraction provides better insight into the structural and chemical details. Figure 2 shows the Rietveld refinement of powder neutron diffraction data from $SrVO_3$ single crystals ground to powder. The reflections have very narrow peak widths and no secondary phases which confirms the phase purity of the single crystals. Further, no structural distortions other than variations in lattice parameter were observed as a function of crystal pulling rate. Final refinement statistics are shown in Table 2. The lattice parameters from neutron diffraction should be interpreted with care, as the time-of-flight technique precludes high accuracy determination of lattice constants, especially for cubic materials. Further, neutron diffraction is rather insensitive to the presence of vanadium ions (the coherent neutron scattering length of vanadium is very small). Thus, the Rietveld refinements are most sensitive to the Sr and O content, and in particular the Sr:O ratio reported separately in Table 3.

In order to determine overall stoichiometries, ICP-OES analysis was used to establish the Sr:V ratios. Trace amounts of Ba (~0.5% by mass) were also observed by ICP-OES in the single crystal samples during this analysis, likely from impurities in the precursor $SrCO_3$. The final determined stoichiometries for the three types of samples studied here are given in Table 3. To the precision of these measurements, growth rate appears to impact the final stoichiometry, but in a non-trivial way.

**Heat capacity**

To evaluate the impact of the compositional changes, heat capacity measurements, which measure the bulk properties of samples, were performed on single crystals grown at 2, 4, and 6 mm/hr. Figure 3 shows specific heat plotted as $C_p/T^3$ versus T. When scaled in this way, phonons that behave according to the Debye model (acoustic phonons and broad bandwidth optical phonons) produce a flat response at low temperatures, followed by a decay towards zero as temperature is increased, whereas highly localized phonons (small bandwidth, optical) follow the Einstein model and present as humps in the specific heat. We immediately find that in addition to the expected Debye phonon contribution that there is also an Einstein contribution at T ~ 30 K. This implies that there is a low-lying, narrow bandwidth (localized) optical phonon at 150 K ~ 15 meV ~ 100 $cm^{-1}$ and these are also seen in other perovskites.[26] Further, there is an upturn at the lowest temperatures indicative of electronic, magnetic, and/or Schottky contributions.

To quantify the behavior of the phonons in $SrVO_3$ and its change in amplitude with composition, we utilized an analytical model combining a Sommerfeld coefficient with a spin fluctuation term (to capture the low temperature electronic portion, vide infra), and Debye, and Einstein specific heat contributions to describe a complex phonon density of states:

$$C_V(T) = C_{el}(\mathrm{T}) + C_D(\mathrm{T}) + C_E(T) \quad (1)$$

$$C_D(\mathrm{T}) = 9s_D\mathrm{R}\left(\frac{T}{\theta_D}\right)^3 \int_0^{\theta_D/T} \frac{(\theta/T)^4 e^{\theta/T}}{[e^{\theta/T}-1]^2}\, \mathrm{d}\frac{\theta}{T} \quad (2)$$

$$C_E(\mathrm{T}) = 3s_E\mathrm{R}\left(\frac{\theta_E}{T}\right)^2 \frac{e^{\theta_E/T}}{[e^{\theta/T}-1]^2} \quad (3)$$

$$C_{el}(\mathrm{T}) = \gamma\mathrm{T} + \mathrm{A}T^3\ln({}^{T}/_{T_{sf}})e^{-{}^{T}/_{T_{sf}}} \quad (4)$$

where $\theta_D$ is the Debye temperature; $\theta_E$ is the Einstein temperature; $s_D$, and $s_E$ are, respectively, the Debye and Einstein oscillator strengths; R is the molar Boltzmann constant; γ is the Sommerfeld coefficient describing an electronic contribution, and A and $T_{sf}$ are the spectral weight and spin fluctuation temperature, respectively. The experimental data at T = 2 – 300 K can be best described by one Einstein and one Debye mode for the phonon contribution. The results of these fits are shown in Table 4.

The total number of oscillators is close to the total number of atoms present in a unit cell of $SrVO_3$ (4.6-4.9) suggesting good agreement between the model and the $SrVO_3$ system. Changes in composition change the intensity of the Einstein contributing peak but not its position, indicating a shift of phonon spectral weight between the Debye and Einstein contributions. One likely explanation of the variation of the intensity of the Einstein mode with small changes in composition is that it is related to changed local dynamics in the presence of greater strontium and/or oxygen vacancies: more vacancies make the perovskite framework less rigid and would allow more local flexing. This is consistent with the observation that the most stoichiometric specimen shows the smallest amplitude of the Einstein mode.

More accurate information can be obtained about the electronic properties from analysis of the low temperature portion of the specific heat. A plot of Cp/T vs. $T^2$ is shown in Figure 4a. The three different single crystal samples show qualitatively and quantitatively distinct behaviors. A pure, simple metal would be expected to show a linear response in such a plot. Deviations from linearity indicate either electron-electron correlations, or electronic scattering/interaction with defects. The data can be quantitatively modeled using the expression:

$$\frac{C_p}{T} = \gamma + \beta^I T^2 + AT^2 ln(T) \quad (5)$$

where $\beta^I = \beta_3 - Aln(T_{sf})$, $\beta_3$ is the phonon contribution, γ is the Sommerfeld coefficient corresponding to the electronic term, A is a scale factor for interactions arising from the metallic electrons with localized states (such as Kondo impurities), and $T_{sf}$ is a characteristic interaction energy scale. Results of fits, yielding γ, $\beta^I$, and A are given in Table 5. Figure 4b shows a plot of the obtained values of $\beta^I$ and A to extract estimated values of $T_{sf}$ = 14.3(5) K and $\beta_3$ = 47.7(5) μJ/mol/$K^4$. The value of $T_{sf}$ obtained in this way is between the values extracted from the high temperature fits. Further, the $\beta_3$ extracted in this way gives an average Debye temperature of 590 K, comparable to those found from the high temperature fits. The values of γ are in good agreement with those found from the fits of the data up to higher temperature. The large variation in γ between samples is unexpected: band structures[27] for $SrVO_3$ show a smooth, slowly varying density of states. Further, the observed magnitudes higher than the calculated density of states $\gamma_{DOS}$ = 3.5 mJ $mol^{-1}$ $K^{-2}$.[i] Additionally, the spin fluctuation contribution also exhibits substantial variations between samples. This suggests an enhanced role of electron correlations (either between band electrons or between band electrons and localized states). Future works can explore in detail the origin of these differences.

**Resistivity**

The electrical resistivity of the $Sr_{0.985(7)}VO_{2.99(2)}$ sample was measured with a Quantum Design PPMS using the standard four-probe configuration to eliminate contact resistance, Figure 5. It increases with temperature as expected for a metal. The data can be fit well to the model:

$$\rho = \rho_0 + AT^2 \quad (6)$$

where $\rho_0$ is the residual resistance and A is a factor describing the strength of electron-electron scattering. The resistivity is consistent with dominant electron-electron scattering all the way up to room temperature, with A = 5.20 nΩcm$K^{-2}$. This implies that the resistivity is not strongly governed by the scattering of phonons, but rather strong electron-electron interactions. This is consistent with the low temperature specific heat (which also indicates strong electron-electron interactions in this material) and implies that electron-phonon interactions are comparatively weak.

Our measured single-crystal residual resistivity ratio (RRR) for the single crystal measured is lower than results achieved by the thin film community but higher than previous single crystal results, Table 6.[28-30] While a high RRR is normally taken as an indicator of sample quality, it is not clear that is an appropriate interpretation here given the dominance of electron-electron scattering. More specifically, if electron-electron correlations reduce as the composition deviates from ideality, then the RRR might grow faster from that than the loss from greater impurity scattering, resulting in a larger RRR with a larger off stoichiometry. Unfortunately, the inability to measure specific heat on thin films precludes a more rigorous comparison and judgement of the value of RRR in assessing materials quality in this instance.

Using the specific heat and resistivity, we can calculate the Kadowaki-Woods ratio, which is a metric for drawing comparisons between the electron-electron scattering and electron mass renormalization between strongly correlated electron materials. Calculated as the ratio of the A/$\gamma^2$, simple metals are found to adopt a nearly universal value 0.4 μΩ.cm.$mol^2$ $K^2$ $J^{-2}$, with values increasing with increasing electron correlations. We find that the KWR of $Sr_{0.985(7)}VO_{2.91(4)}$ is A/$\gamma^2$ = 11 μΩ cm $mol^2$ $K^2$ $J^{-2}$. This value is above transition metals and comparable to heavy fermion systems. It is also comparable to the value of 10 μΩ cm $mol^2$ $K^2$ $J^{-2}$ previously reported for $SrVO_3$.[31] Overall, this implies that $SrVO_3$ is a metal that in single crystal form shows signs of significant electron-electron interactions. This is consistent with calculations[5] that suggest e-e correlations drive a band narrowing versus a nearly free electron picture. This conclusion is also supported by other experimental results such as angle-resolved photoemission spectroscopy on single crystals of $SrVO_3$ lending further support to the claim that this material is more than just a canonical metal.[32-33]

---

[i] $[\gamma_{DOS} = \pi^{\frac{2}{3}} {k_B}^2 g(\varepsilon_F) = \pi^{\frac{2}{3}} {k_B}^2 * 8.62 * 10^{-5} eV.K^{-1} * 1.5\ states.eV^{-1}.cell^{-1} * N_A\ cell.mol^{-1} = 3.5$ mJ $mol^{-1}$ $K^{-2}$.]

## Conclusion

Single crystals of nearly stoichiometric $SrVO_3$ were successfully produced from polycrystalline $SrVO_3$ using the laser diode floating zone. Using a ratio of 85:15 $SrVO_3$ to its thermodynamically stable intermediate $Sr_2V_2O_7$ was found to aid growth. The slowest single-crystal growth rate, 2 mm/hr, produced $SrVO_3$ with the smallest lattice parameter and a composition of $Sr_{0.985(7)}VO_{2.91(4)}$. This value is equal to that achieved in polycrystalline syntheses. The *RRR* value achieved in this work is higher than previous reports for single-crystalline $SrVO_3$, although our work casts doubt on *RRR* being an adequate measure of crystalline quality in this case. It also suggests that this material would be a good candidate for thin film growth targets to produce films of complex oxides for electronic applications in addition to its known applications in catalysis and thermionic emitters.

## Acknowledgements

This work was funded by the Platform for the Accelerated Realization, Analysis, and Discovery of Interface Materials (PARADIM), a National Science Foundation Materials Innovation Platform (NSF DMR-1539918). The authors would like to thank M. Siegler, M.J. Winiarski, V.J. Stewart, X.(J.) Zhang, M. Sinha, A. Volzke, and T. Tran for technical assistance. A portion of this research used resources at the Spallation Neutron Source, a DOE Office of Science User Facility operated by the Oak Ridge National Laboratory.

## References

[1] Bhalla, A.S.; Guo, R.; and Roy, R. "The perovskite structure—a review of its role in ceramic science and technology," *Mat. Res. Innovat.* **2000**, 4 (1), 3-26. DOI: 10.1007/s100190000062

[2] Megaw, H.D. "Crystal structure of double oxides of the perovskite type," *Proc. Phys. Soc.* **1946**, 58 (2), 133-152. DOI: 10.1088/0959-5309/58/2/301

[3] Jena, A.K.; Kulkarni, A.; and Miyasaka, T. "Halide Perovskite Photovoltaics: Background, Status, and Future Prospects," *Chem. Rev.* **2019**, 119, 3036-3103. DOI: 10.1021/acs.chemrev.8b00539

[4] Filip, M.R. and Giustino, F. "The geometric blueprint of perovskites," *Proceedings of the National Academy of Sciences* **2018**, 115 (21), 5397-5402. DOI: 10.1073/pnas.1719179115

[5] Zhang, L.; Zhou, Y.; Guo, L.; Zhao, W.; Barnes, A.; Zhang, H.-T.; Eaton, C.; Zheng, Y.; Brahlek, M.; Haneef, H.F.; Podraza, N.J.; Chan, M.H.W.; Gopalan, V.; Rabe, Ka.M.; and Engle-Herbert, R. "Correlated metals as transparent conductors," *Nature Materials* **2016**, 15, 204-210. DOI: 10.1038/nmat4493

[6] Shoham, L.; Baskin, M.; Han, M.-G.; Zhu, Y.; and Kornblum, L. "Scalable Synthesis of the Transparent Conductive Oxide $SrVO_3$," *Ad. Electron. Mater.* **2020**, 6 (1), 1900584. DOI: 10.1002/aelm.201900584

[7] Hui, S.; Petric, A.; and Gong, W. "Stability and Conductivity of Perovskite Oxides Under Anodic Conditions," *ECS Proc. Vol.* **1999**, 1999-19, 632. DOI: 10.1149/199919.0632PV

[8] Trikalitis, P.N. and Pomonis, P.J. "Catalytic activity and selectivity of perovskites $La_{1-x}Sr_xV_{1-x}{}^{3+}V_x{}^{4+}O_3$ for the transformation of isopropanol," *Applied Catalysis A: General* **1995**, 131 (2), 309-322. DOI: 10.1016/0926-860X(95)00121-2

[9] Jacobs, R.; Booske, J.; and Morgan, D. "Understanding and Controlling the Work Function of Perovskite Oxides Using Density Functional Theory," *Adv. Funct. Mater.* **2016**, 26, 5471-5482. DOI: 10.1002/adfm.201600243

[10] Jacobs, R.; Morgan, D.; and Booske, J. "High-throughput computational screening for low work function perovskite electron emitters," *2017 Eighteenth International Vacuum Electronics Conference* **2017**, 1-2. DOI: 10.1109/IVEC.2017.8289723

[11] Jacobs, R.; Lin, L.; Ma, T.; Lu-Steffes, O.; Vlahos, V.; Morgan, D.; and Booske, J. "Perovskite electron emitters: Computational prediction and preliminary experimental assessment of novel low work function cathodes," *2018 IEEE International Vacuum Electronics Conference* **2018**, 37-38. DOI: 10.1109/IVEC.2018.8391540

[12] Wang, C.; Zhang, H.; Deepak, K.; Chen, C.; Fouchet, A.; Duan, J.; Hilliard, D.; Kentsch, U.; Chen, D.; Zeng, M.; Gao, X.; Zeng, Y.-J.; Helm, M.; Prellier, W.; and Zhou, S. "Tuning the metal-insulator transition in epitaxial $SrVO_3$ films by uniaxial strain," *Phys. Rev. Mat.*, **2019**, 3, 115001. DOI: 10.1103/PhysRevMaterials.3.115001

[13] Pena, J. I.; Merino, R. I.; Harlan, N. R.; Larrea, A.; de la Fuente, G. F.; and Orera, V.M., "Microstructure of $Y_2O_3$ Doped $Al_2O_3$–$ZrO_2$ Eutectics Grown by the Laser Floating Zone Method," *J. Eur. Ceram. Soc.* **2002**, 22, 2595–2602. DOI: 10.1016/S0955-2219(02)00121-8

[14] Sinha, M.; Pearson, T. J.; Reeder, T. R.; Vivanco, H. K.; Freedman, D. E.; Phelan, W. A.; and McQueen, T. M. “Introduction of Spin Centers in Single Crystals of $Ba_2CaWO_{6-\delta}$,” *Phys. Rev. Materials* **2019**, 3, 125002. DOI: 10.1103/PhysRevMaterials.3.125002.
[15] Berry, T.; Pressley, L.A.; Phelan, W.A.; Tran, T.T.; and McQueen, T. “Laser-Enhanced Single Crystal Growth of Non-Symmorphic Materials: Applications to an Eight-Fold Fermion Candidate,” *Chem. Mater.* **2020**, *23*, 5827. DOI: 10.1021/acs.chemmater.0c01721
[16] Pressley, L. A.; Torrejon, A.; Phelan, W. A.; and McQueen, T. M. "Discovery and Single Crystal Growth of High Entropy Pyrochlores," *Inorg. Chem.* **2020**, 59, 17251-17258. DOI: 10.1021/acs.inorgchem.0c02479
[17] Chandrashekhar, M. V. S.; Letton, J. A.; Williams, T.; Ajilore, A.; and Spencer, M. “Laser-Induced Graphitization of Boron Carbide in Air,” U.S. Patent No. US20190352234-A1, **2019**.
[18] Scudder, M. R.; He, B.; Wang, Y.; Rai, A.; Cahill, D.; Windl, W.; Heremans, J. P.; and Goldberger, J. "Highly Efficient Transverse Thermoelectric Devices with Re4Si7 crystals," *Energy Environ. Sci.* **2021**, 14, 4009-4017. DOI: 10.1039/D1EE00923K
[19] Ogawa, T.; Urata, Y.; Wada, S.; Onodera, K.; Machida, H.; Sagae, H.; Higuchi, M.; and Kodaira, K. “Efficient laser performance of Nd:$GdVO_4$ crystals grown by the floating zone method,” *Opt. Lett*. **2003**, 28, 2333-2335. DOI: 10.1364/OL.28.002333
[20] Carrasco, M. F.; Silva, R. A.; Silva, R. F.; Amaral, V. S.; and Costa, F. M. “Critical Current Density Improvement in BSCCO Superconductors by Application of an Electric Current during Laser Floating Zone Growth.” Physica C **2007**, 460−462, 1347−1348. DOI: 10.1016/j.physc.2007.04.019
[21] Arpino, K.E.; Trump, B.A.; Scheie, A.O.; McQueen, T.M.; and Koohpayeh, S.M. “Impact of Stoichiometry of $Yb_2Ti_2O_7$ on Its Physical Properties,” *Phys. Rev. B* **2017**, 95, 094407 DOI: 10.1103/PhysRevB.95.094407
[22] Sheldrick, G. M. “Acta Crystallogr., Sect. A: Found,” *Crystallogr.* **2008**, 64, 112-122. DOI: 10.1107/S0108767307043930
[23] Farrugia, L. “WinGX and ORTEP for Windows: an Update,” *J. Appl. Crystallogr.* 2012, 45, 849−854.
[24] Toby, B.H. “EXPGUI, a graphical user interface for GSAS,” *J. Appl. Cryst.* **2001**, 34, 210–213. DOI: /10.1107/S0021889801002242
[25] Ardila, D.R.; Andreeta, J.P.; and Basso, H.C. “Preparation, microstructural and electrical characterization of $SrVO_3$ single crystal fiber,” *J. Cryst. Growth* **2000**, 211, 313-317.
[26] Tran, T. T.; Panella, J. R.; Chamorro, J. R.; Morey, J. R.; and McQueen, T. M. “Designing indirect–direct bandgap transitions in double perovskites,” *Mater. Horiz*. **2017**, 4, 688−693. DOI: 10.1039/C7MH00239D
[27] Ishida, H.; Wortmann, D.; and Liebsch, A. “Electronic structure of SrVO3(001) surfaces: A local-density approximation plus dynamical mean-field theory calculation,” *Phys. Rev. B* **2006** 73, 245421. DOI: 10.1103/PhysRevB.73.245421
[28] Lan, Y. “Structure, Magnetic Susceptibility and Resistivity Properties of $SrVO_3$,” *Journal of Alloys and Compounds* **2003,** 354 (1-2), 95-98. DOI: 10.1016/S0925-8388(02)01349-X
[29] Onoda, M.; Ohta, H.; and Nagasawa, H. “Metallic properties of perovskite oxide $SrVO_3$,” *Solid State Communication* **1991**, 79, 281-285. DOI: 10.1016/0038-1098(91)90546-8
[30] Bérini, B.; Demange, V.; Bouttemy, M.; Popova, E.; Keller, N.; Dumont, Y.; and Fouchet, A. “Control of High Quality $SrVO_3$ Electrode in Oxidizing Atmosphere,” *Adv. Mater. Interfaces* **2016**, 3, 1600274. DOI: 10.1002/admi.201600274
[31] Inoue, I. H.; Goto, O.; Makino, H.; Hussey, N. E.; and Ishikawa, M. “Bandwidth control in a perovskite-type $3d^1$-correlated metal $Ca_{1-x}Sr_xVO_3$. I. Evolution of the electronic properties and effective mass,” *Phys. Rev. B*. **1998**, 58, 4372. DOI: 10.1103/PhysRevB.58.4372
[32] Moyer, J.A.; Eaton, C.; and Engel-Herbert, R. “Highly Conductive $SrVO_3$ as a Bottom Electrode for Functional Perovskite Oxides,” *Adv. Mater.* **2013**, 25, 3568-3582. DOI: 10.1002/adma.201300900
[33] Georges, A.; Kotliar, G.; Krauth, W.; and Rozenberg, M.J. “Dynamical mean-field theory of strongly correlated fermion systems and the limit of infinite dimensions,” *Rev. Mod. Phys.* **1996**, 68 (1). DOI: 10.1103/RevModPhys.68.13

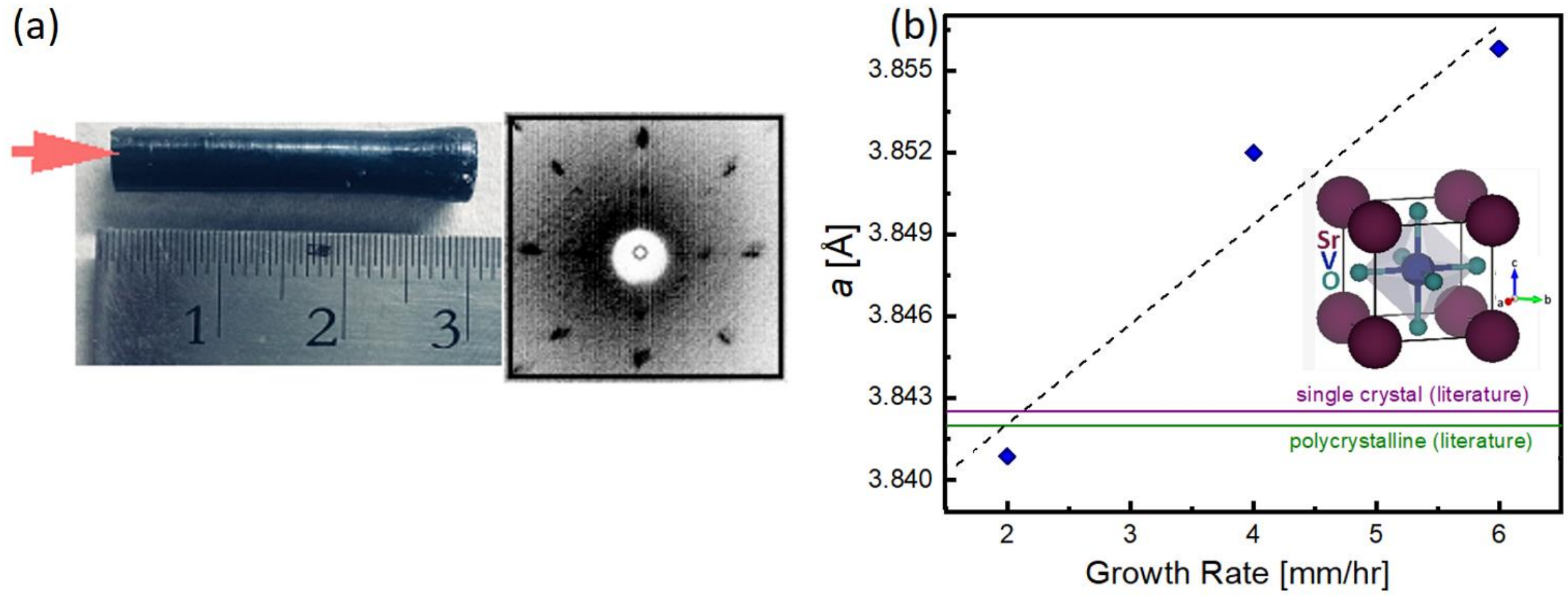

**Figure 1**. (**a)** Photograph of a large single crystal of $SrVO_3$ and a Laue diffraction pattern of the [110] direction along the growth direction as indicated with the coral colored arrow. **(b)** A pseudo-Vegard's plot showing change in measured lattice parameter with growth rate and a unit cell of $SrVO_3$, measured on a laboratory powder diffractometer with a Si internal standard The blue diamonds represent the three single crystal samples grown in this work; the green and purple lines are literature polycrystalline and single crystal samples.[18,20] The error bars of the lattice parameter are within the range of the data points.

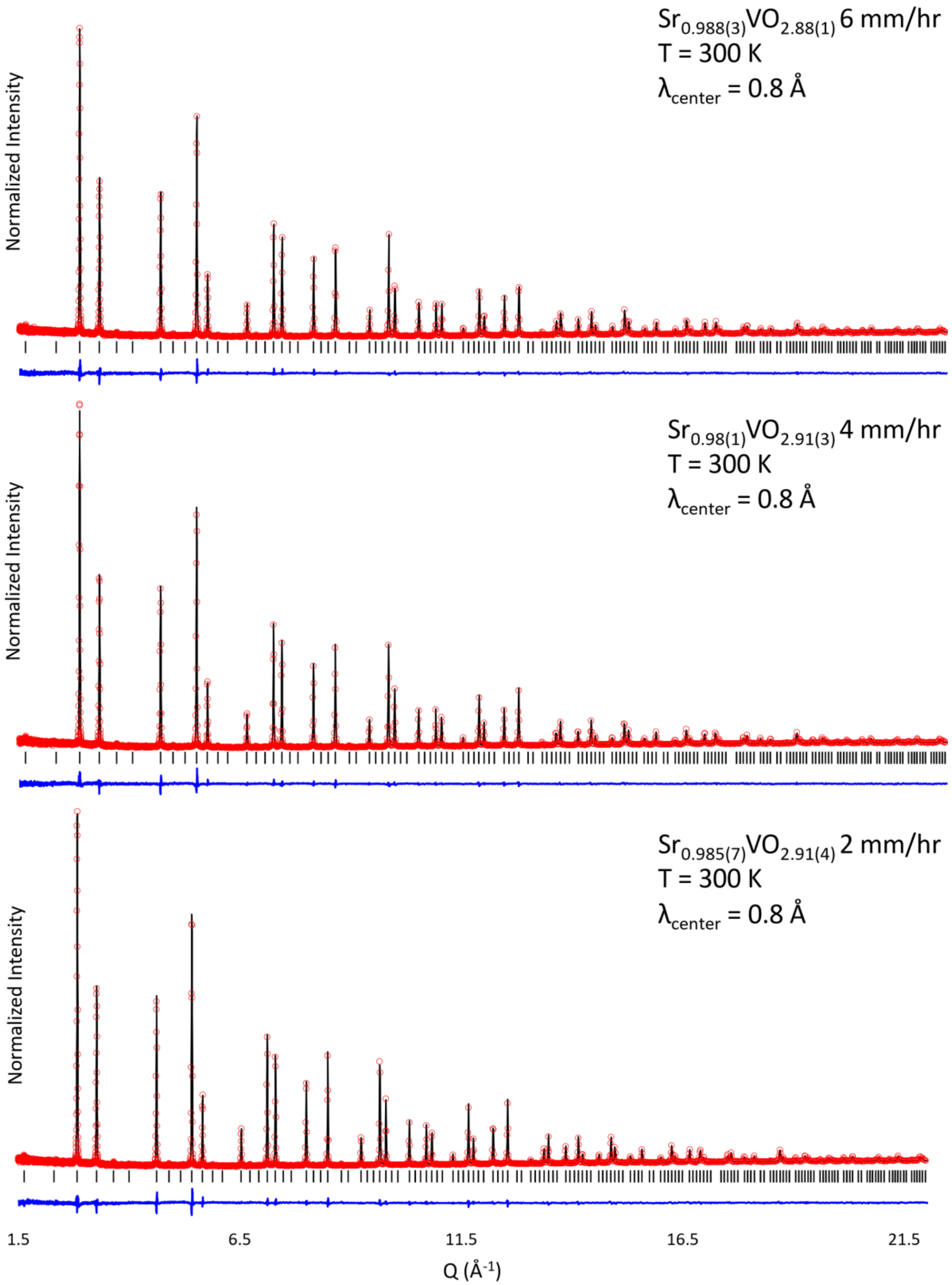

**Figure 2.** Rietveld refinement (black curve) of POWGEN neutron diffraction data (red circles) for three single crystal $SrVO_3$ samples ground to powders. The difference curve between fit and data is the blue line. Black tick marks are shown at the expected reflections. Axes are the same for all three fits.

| Formula | $SrVO_3$ |
|---|---|
| Crystal system | Cubic |
| Space Group | *Pm*-3*m* (No. 221) |
| a (Å) | 3.844(5) |
| V ($Å^3$) | 56.8 (1) |
| Z | 2 |
| M/$gmol^{-1}$ | 186.56 |
| $\rho_0/gcm^{-3}$ | 10.908 |
| $\mu/mm^{-1}$ | 54.568 |
| Radiation | Mo K$\alpha$, λ= 0.71073 Å |
| Temperature (K) | 293(2) K |
| Adsorption Coefficient | 44.56 |
| Reflections collected/number of parameters | 1430/38 |
| $R_{int}$[a] | 0.035 |
| Goodness-of-fit | 1.275 |
| R[F][b] | 0.0077 |
| $R_w(F_0^2)$[c] | 0.0223 |

[a] $\boldsymbol{R_{int}} = \sum|\boldsymbol{F_0^2} - \boldsymbol{F_0^2}(\boldsymbol{mean})| / \sum|\boldsymbol{F_0^2}|$ [b] $\boldsymbol{R_{int}} = \sum||\boldsymbol{F_0}| - |\boldsymbol{F_c}|| / \sum|\boldsymbol{F_0}|$

[c] $\boldsymbol{Rw}(\boldsymbol{F_0^2}) \sqrt{\sum(\boldsymbol{w}(\boldsymbol{F_0^2} - \boldsymbol{F_c^2})^2) / \sum(\boldsymbol{w}(\boldsymbol{F_0^2})^2)}$,

$\boldsymbol{w} = \boldsymbol{1}/(\boldsymbol{\sigma^2}(\boldsymbol{F_0^2}) + (\boldsymbol{a}((\boldsymbol{2F_c^2} + \boldsymbol{Max}(\boldsymbol{F_0^2}, \boldsymbol{0}))/\boldsymbol{3}))^{\boldsymbol{2}})$, **a = 0.0163**

| | Occ. | Wyckoff Positions | x (Å) | y (Å) | z (Å) | $U_{iso}$ ($Å^2$) |
|---|---|---|---|---|---|---|
| Sr | 1 | 1a | 0.0 | 0.0 | 0.0 | 0.00970 |
| V | 1 | 1b | 0.5 | 0.5 | 0.5 | 0.00715 |
| O | 1 | 3c | 0.5 | 0.0 | 0.5 | 0.01318 |

**Table 1.** Crystallographic parameters of the SXRD for $Sr_{0.985(7)}VO_{2.91(4)}$ or equivalently the "2mm/hr" growth.

| $Sr_{0.988(3)}VO_{2.88(1)}$ 6 mm/hr, a = 3.845978(9) Å<br>$\chi^2$ = 3.17 $R_{wp}$ = 4.68% | | | | | | |
|---|---|---|---|---|---|---|
| **Atom** | **Wycoff symbol** | **x (Å)** | **y (Å)** | **z (Å)** | **Occupancy** | **$U_{iso}$ ($Å^2$)** |
| Sr | 1b | 0.5 | 0.5 | 0.5 | 1 | 0.00529(23) |
| V | 1a | 0.0 | 0.0 | 0.0 | 1 | 0.0014(5) |
| O | 3d | 0.5 | 0.0 | 0.0 | 0.970(9) | 0.00477(9) |
| **$Sr_{0.98(1)}VO_{2.91(3)}$ 4 mm/hr, a = 3.846638(10) Å**<br>**$\chi^2$ = 2.82 $R_{wp}$ = 4.45%** | | | | | | |
| **Atom** | **Wycoff symbol** | **x (Å)** | **y (Å)** | **z (Å)** | **Occupancy** | **$U_{iso}$ ($Å^2$)** |
| Sr | 1b | 0.5 | 0.5 | 0.5 | 1 | 0.00428(17) |
| V | 1a | 0.0 | 0.0 | 0.0 | 1 | 0.0032(6) |
| O | 3d | 0.5 | 0.0 | 0.0 | 0.991(9) | 0.00531(7) |
| **$Sr_{0.985(7)}VO_{2.91(4)}$ 2 mm/hr, a = 3.844644(9) Å**<br>**$\chi^2$ = 2.48 $R_{wp}$ = 4.24%** | | | | | | |
| **Atom** | **Wycoff symbol** | **x (Å)** | **y (Å)** | **z (Å)** | **Occupancy** | **$U_{iso}$ ($Å^2$)** |
| Sr | 1b | 0.5 | 0.5 | 0.5 | 1 | 0.00440(17) |
| V | 1a | 0.0 | 0.0 | 0.0 | 1 | 0.0027(5) |
| O | 3d | 0.5 | 0.0 | 0.0 | 0.982(9) | 0.00513(7) |

**Table 2.** Residuals and crystallographic parameters obtained from fits to the neutron diffraction data for ground single crystals from three different growth rates. $^{A}\chi^2 = \frac{\sum(w(I_0^2-I_c^2)^2)}{(N_{obs}-N_{var})}$ $^{b}\boldsymbol{Rwp} = \sqrt{\sum(\boldsymbol{w}(\boldsymbol{I_0^2}-\boldsymbol{I_c^2})^2)/\sum(\boldsymbol{w}(\boldsymbol{I_0^2})^2)}$

| Sample | Sr:V (ICP) | O:Sr (POWGEN) | Overall stoichiometry | Notes |
|---|---|---|---|---|
| 6 mm/hr | 0.988(3) | 2.91(3) | $Sr_{0.988(3)}VO_{2.88(1)}$ | Contained 0.5 wt% Ba (ICP) |
| 4 mm/hr | 0.98(1) | 2.97(3) | $Sr_{0.98(1)}VO_{2.91(3)}$ | |
| 2 mm/hr | 0.985(7) | 2.95(3) | $Sr_{0.985(7)}VO_{2.91(4)}$ | |

**Table 3.** Comparison of $SrVO_3$ stoichiometry as a function of growth rate (in mm/hr). Composition was determined using a combination of two techniques. The Sr:V ratios were determined using inductively coupled plasma optical emission spectrometry (ICP-OES). Since O is not a good parameter to measure using the ICP method, the O:Sr ratios were determined from neutron diffraction results, in which V does not appear due to its low coherent scattering length. In each case, specimens were taken from the same lengthwise portion of the boule.

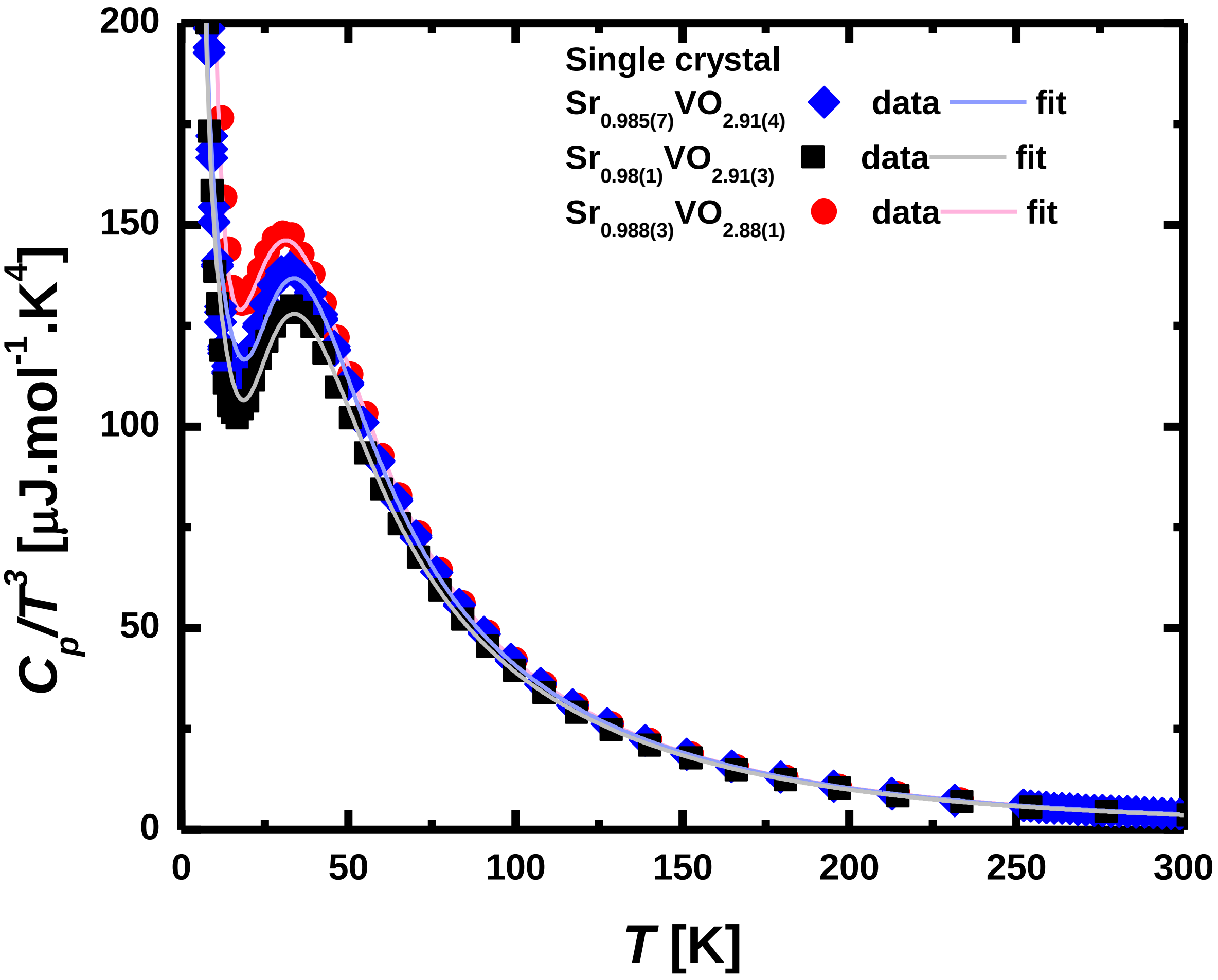

**Figure 3.** The heat capacity over temperature cubed versus temperature demonstrates an Einstein-like mode at *ca.* 30 K. The lowest intensity Einstein mode was observed in the sample with the slowest growth rate ($Sr_{0.985(7)}VO_{2.99(2)}$). The error bars are within the data points.

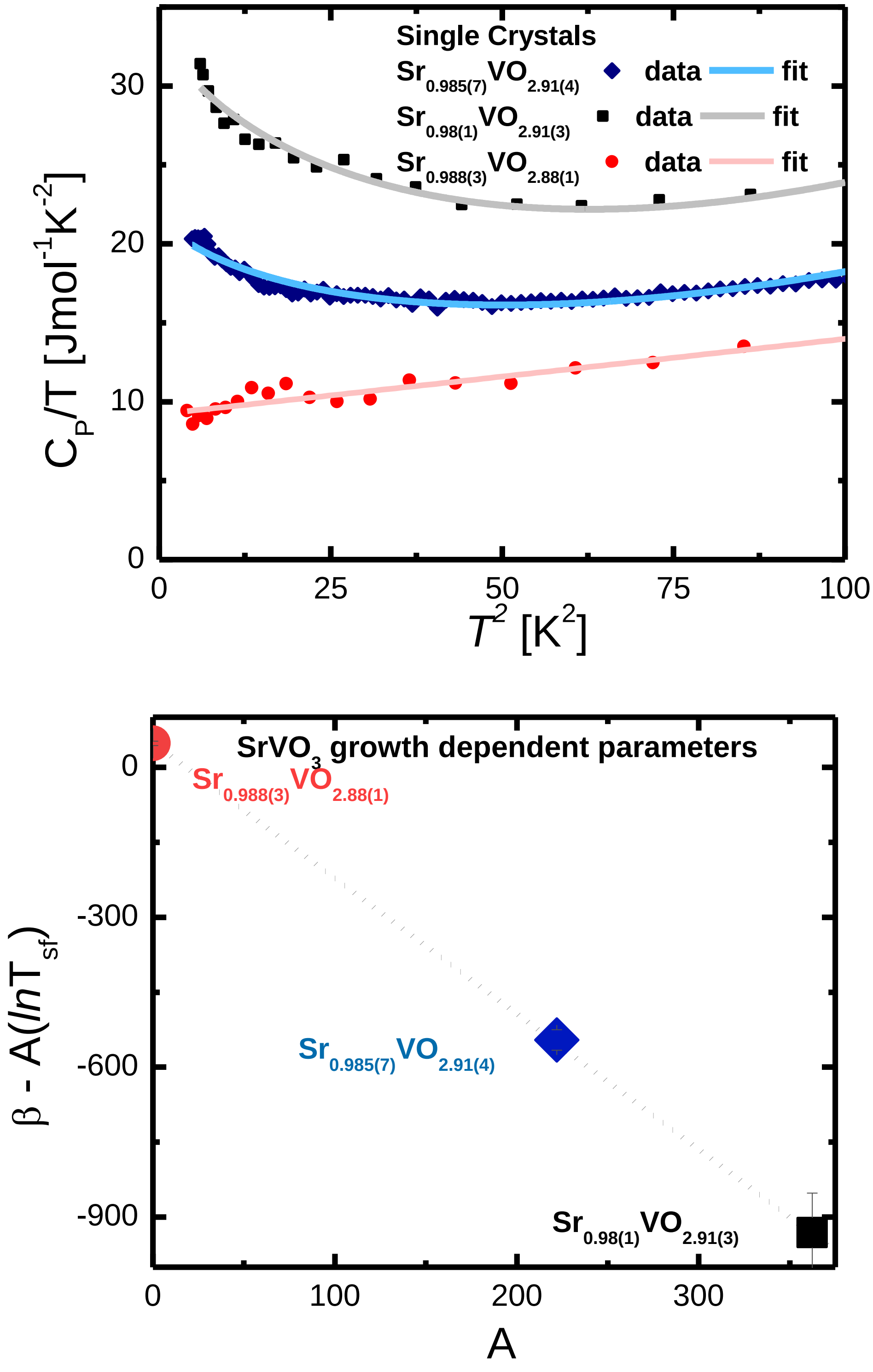

**Figure 4.** (a) Heat Capacity over temperature versus temperature squared. Note: the data was collected on the specimens on top of the crystal (closer to the molten zone). The data as a function of composition is modeled to the equation: $\frac{C_p}{T} = \gamma + \beta^I T^2 + AT^2 ln(T)$, where $\beta^I = \beta_3 - Aln(T_{sf})$ . (b) The y-intercept extrapolated from the plot of $\beta^1$ vs $A$ gives the phonon contribution, and the slope gives an estimate of the effective spin fluctuation temperature. The extracted values are $\beta_3 = 47.7(5)$ μJ/mol-K$^{-4}$ and $T_{sf} = 14.3(5)$ K.

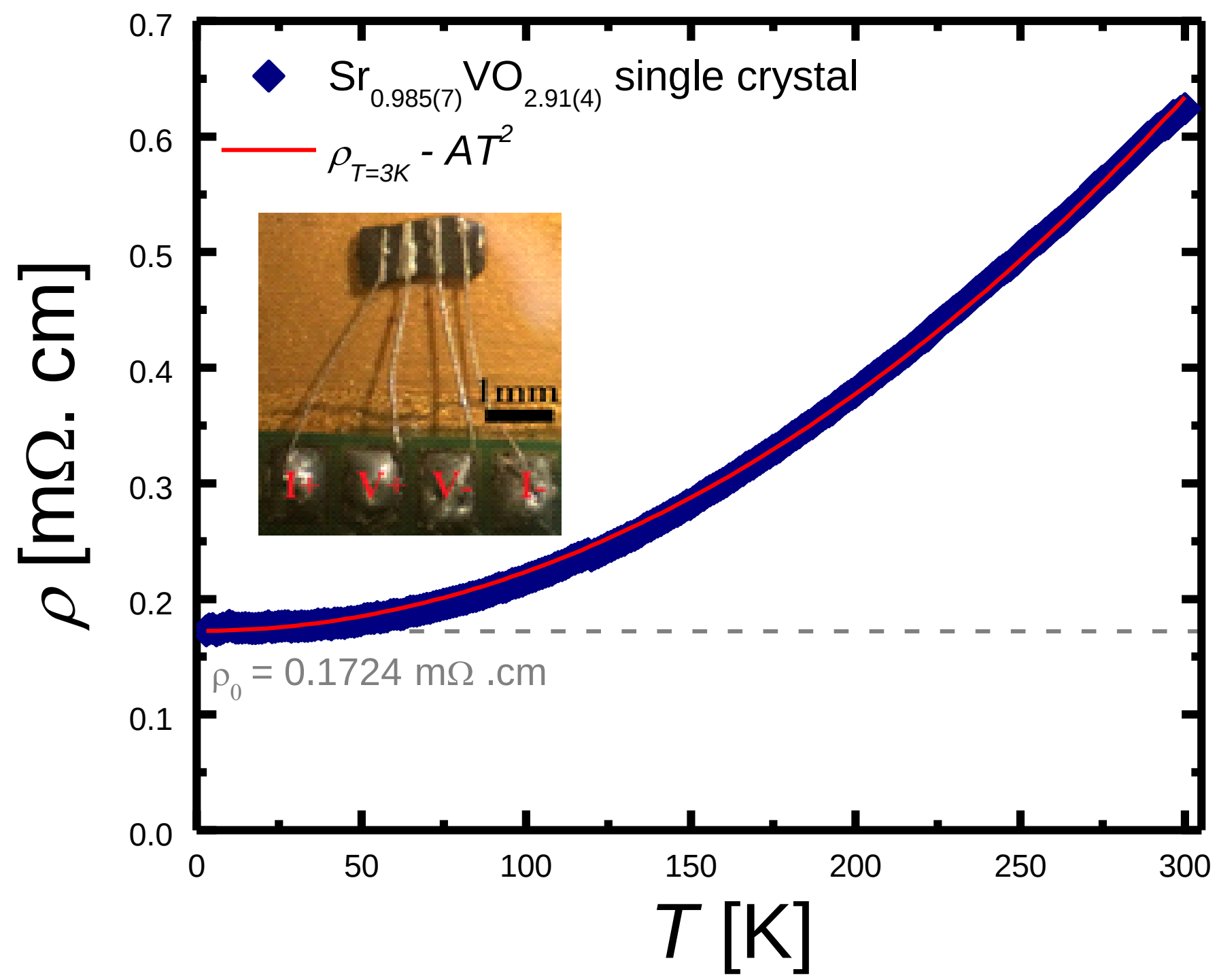

**Figure 5**. Temperature dependency of resistivity for the $Sr_{0.985(7)}VO_{2.99(2)}$ sample, believed to be the purest single-crystalline sample. The residual resistivity at T = 3 K is 0.1724 mΩ-cm. The fit line is a model using the residual resistivity at T=3 K and an electron-electron scattering term (see text).

| Sample composition | $s_D$ (oscillator strength/formula unit) | $s_E$ (oscillator strength/formula unit) | γ (mJ/mol-$K^{-2}$) | $θ_D$(K) | $θ_E$(K) | $T_{sf}$ | A (J.mol$^{-1}$K$^{-4}$) |
|---|---|---|---|---|---|---|---|
| $Sr_{0.988(3)}VO_{2.88(1)}$ | 3.6(1) | 0.96(2) | 12(1) | 665(6) | 174(1) | 0 | 0 |
| $Sr_{0.98(1)}VO_{2.91(3)}$ | 4.0(1) | 0.60(4) | 37(1) | 650(4) | 191(4) | 20.1(3) | 4.2(2) x $10^{-4}$ |
| $Sr_{0.985(7)}VO_{2.91(4)}$ | 3.6(1) | 1.30(2) | 19(1) | 727(4) | 195(2) | 10.1(5) | 4.4(2) x $10^{-4}$ |

**Table 4.** Fitting Parameters to the $Cp/T^3$ Data for three single crystalline samples.

| Sample composition | γ (mJ/mol-K$^{-2}$) | $\beta^{I}$ | $\beta_3$ (μJ/mol-K$^{-4}$) | α (μJ/mol-K$^{-2}$) |
|---|---|---|---|---|
| $Sr_{0.988(3)}VO_{2.88(1)}$ | 9.21(2) | 0 | 47.7(4) | 0 |
| $Sr_{0.98(1)}VO_{2.91(3)}$ | 33.55(5) | -930(79) | 47.7(5) | 362(33) |
| $Sr_{0.985(7)}VO_{2.91(4)}$ | 21.71(2) | -545(20) | 47.7(5) | 221(8) |

**Table 5.** Fitting Parameters to the Cp/T$^2$ Data for three single crystalline samples.

| **Type of material** | **Temperature range (K)** | ***RRR*** | **References** |
|---|---|---|---|
| Polycrystalline | 2-300 | 10.0 | 24 |
| Thin Films | 5-300 | ~222 | 1 |
| Single crystal | 4.5-300 | 1.50 | 23 |
| **Single crystal** | **3-300** | **3.62** | **This work** |

**Table 6.** Comparison of the highest $SrVO_3$ residual resistivity ratio values on polycrystalline, thin films, and bulk single crystal materials including this work on the $Sr_{0.985(7)}VO_{2.99(2)}$ single crystals.